\documentclass[11pt,preprint]{aastex}
\usepackage{epsfig,color}
\usepackage{mathrsfs}
\usepackage{graphicx}
\usepackage{bm}
\usepackage{amsmath}


\shorttitle{Short Term Variability of S5 0716+714} \shortauthors{Feng et al.}

\begin{document}

\title{Spectroscopic Monitoring of Blazar S5 0716+714: Brightness-Dependent Spectral Behavior}

\author{Hai-Cheng Feng\altaffilmark{1,2,3, \bigstar},
Sen. Yang\altaffilmark{2,4},
Zi-Xu. Yang\altaffilmark{2,4},
H. T. Liu\altaffilmark{1,3,5, \bigstar},
J. M. Bai\altaffilmark{1,3,5, \bigstar},
Sha-Sha. Li\altaffilmark{2,4},
X. H. Zhao\altaffilmark{1,3,5},
Jin. Zhang\altaffilmark{6},
Y. B. Li\altaffilmark{1,2,3},
M. Xiao\altaffilmark{4},
Y. X. Xin\altaffilmark{1,3,5},
L. F. Xing\altaffilmark{1,3,5},
K. X. Lu\altaffilmark{1,3,5},
L. Xu\altaffilmark{1,3,5},
J. G. Wang\altaffilmark{1,3,5},
C. J. Wang\altaffilmark{1,3,5},
X. L. Zhang\altaffilmark{1,3,5},
J. J. Zhang\altaffilmark{1,3,5},
B. L. Lun\altaffilmark{1,3,5},
S. S. He\altaffilmark{1,3,5}}

\altaffiltext{1} {Yunnan Observatories, Chinese Academy of Sciences,
Kunming 650011, Yunnan, People's Republic of China}

\altaffiltext{2} {University of Chinese Academy of Sciences, Beijing
100049, People's Republic of China}

\altaffiltext{3} {Key Laboratory for the Structure and Evolution of
Celestial Objects, Chinese Academy of Sciences, Kunming 650011,
Yunnan, People's Republic of China}

\altaffiltext{4} {Key Laboratory for Particle Astrophysics, Institute of High Energy Physics, Chinese Academy of Sciences, 19B Yuquan Road, Beijing, 100049, People's Republic of China}

\altaffiltext{5} {Center for Astronomical Mega-Science, Chinese Academy of Sciences, 20A Datun Road, Chaoyang District, Beijing, 100012, People's Republic of China}

\altaffiltext{6} {Key Laboratory of Space Astronomy and Technology, National Astronomical Observatories, Chinese Academy of Sciences, Beijing 100012, People's Republic of China}

\altaffiltext{$^{\bigstar}$}{Corresponding authors: Hai-Cheng. Feng, e-mail: hcfeng@ynao.ac.cn,
H. T. Liu, e-mail: htliu@ynao.ac.cn, J. M. Bai, e-mail: baijinming@ynao.ac.cn}

\begin{abstract}
  In this paper, we report the new results of spectroscopic observations of $\gamma$-ray blazar S5 0716+714 from 2019 September to 2020 March with the 2.4 m optical telescope at Lijiang Observatory of Yunnan Observatories. The median cadence of observations is $\sim$ 1 day. During the second observation period (Epoch2), the observational data reveal an extremely bright state and a bluer-when-brighter (BWB) chromatism. The BWB trend of Epoch2 differs significantly from that of the first observation period (Epoch1). A significantly brightness-dependent BWB chromatism emerges in the total data of Epoch1 and Epoch2. The BWB trend becomes weaker towards the brighter states, and likely becomes saturated at the highest state. Based on a log-parabolic function, a power-law of synchrotron peak flux and frequency $\nu_{\rm{p}}$, and a power-law of the curvature of synchrotron spectrum and its $\nu_{\rm{p}}$, simulation well reproduces the brightness-dependent BWB trend of S5 0716+714. The BWB trend is seemingly controlled by the shift of $\nu_{\rm{p}}$ with respect to the observational window, and effectively may be dominated by the variations of electron average energy and magnetic field in emitting region.

\end{abstract}

\keywords{Active galactic nuclei (16); BL Lacertae objects (158); Blazars (164); Relativistic jets (1390); Spectroscopy (1558)}

\section{Introduction}
Blazars belong to a subclass of active galactic nuclei including BL Lacertae objects (BL Lacs) and flat-spectrum radio quasars (FSRQs) \citep[e.g.,][]{GT15}. BL Lacs are characterized by featureless optical spectra, rapid variability across the entire electromagnetic spectrum, strong polarization and nonthermal radiation, which are generated by relativistic jets with small viewing angles \citep[e.g.,][]{UP95}. In the $\log(\nu f_{\nu}$) vs. $\log \nu$ plots, their spectral energy distributions (SEDs) show two peaks located at infrared (IR) to X-ray and $\gamma$-ray. The peak at low frequencies can be interpreted as the synchrotron radiation of relativistic electrons in the ralativistic jets, while the $\gamma$-ray peak may be generated from the inverse Compton (IC) scattering of soft photons by the relativistic electrons \citep[e.g.,][]{MG85, Gh98, Ta98, Zh13}.

Flux variations at different timescales are usually accompanied by changes of spectral indices, which provide a useful probe for physical processes in jets of BL Lacs. However, even for a certain source, a few broadband photometric observations show different results which favor several different competing models \citep[e.g.,][]{St06,Da15,Ag16,Ka18}. Therefore, the relationship between spectral behavior and brightness (hereafter index-flux) is still a debated issue. The bluer-when-brighter (BWB) chromatic trends are one of the most common phenomena in BL Lacs, and can be regarded as the evidence of a shock-in-jet model \citep[e.g,][]{MG85}. Although, other spectral behaviors in some sources can be explained by accretion disk based models \citep[e.g.,][]{Ra07} or contamination of the host galaxies \citep{Ni08}, the jet--dominated and weak host galaxy sources are still challenges of our conception of jets.

S5 0716+714 is a typical BL Lac object with extreme variability, prominent jet component, and negligible host contribution. It is therefore an ideal candidate for investigating the properties of jets. The spectroscopic observations of the source in \citet{Fe20} indicate that the BWB trends may depend on the brightness, i.e., the correlations at the bright state are weaker than those at the faint state. However, only 47 data points with $\sim$ 1.7 mag brightness variation can not give a reliable conclusion. Similarly, \citet{Ag16} did not find any correlation between magnitude and colors at the brightest state of the source, while other observations showed strong BWB trends during the faint state \citep[e.g.,][]{Da15}. These observations only base on several broadband photometry, and the bandwidth might affect the correlation coefficient (see details in F20). Moreover, most observations are usually run with low sampling rates, and show small brightness variations ($<$1.5 mag). The brightness-dependent BWB trends are rarely reported in previous studies. The observations with high cadence and large variability amplitude are therefore important for our understanding of the physical conditions in jets.

In this paper, we present the spectroscopic monitoring results of S5 0716+714, which exhibit a significantly brightness-dependent BWB trend. The median cadence of observations is $\sim$ 1 day. Compared to the data in F20, the brightness is increased by a large factor of 13.63 ($\sim$ 2.85 mag). Therefore, we are able to investigate the detailed properties of jets. In Section 2, we introduce the information of observations and data reduction. We present the results and our analysis in Section 3. In Section 4, we discuss the possible radiation processes in the source and give conclusions.

\section{Observations and Data Reduction}
The spectroscopic observations of S5 0716+714 are almost the same as those described in F20 except for minor differences in flux calibration. All the data were taken by the Yunnan Faint Object Spectrograph and Camera, mounted at Cassegrain focus of 2.4 optical telescope at Lijiang Observatory (Yunnan Observatories, Chinese Academy of Science) from 2019 September to 2020 March. The monitoring campaign spanned 166 days, and 106 spectra were obtained in 102 nights. A wide slit with 5$^{\prime\prime}_{\cdot}$05 was adopted to minimize the effects of seeing. We used Grism 3 resulting a wide wavelength coverage 3400--9100 \AA\ and a dispersion of 2.93 \AA\ pixel$^{-1}$. Besides, a UV-blocking filter was used and it can well eliminate the contamination of the second-order spectrum. The final effective spectra cover 4250--8050 \AA. In each night, we simultaneously put the object and a comparison star in the slit, and this can give high-precision flux calibration. Since a high quality spectrum of comparison star has been obtained in F20, we do not observe spectrophotometric standard star in this observing season. All the spectra are calibrated using the same spectrum of comparison star in F20.

The raw data are reduced with standard IRAF routines. The spectra were extracted with a aperture radius of 21 pixels (5$^{\prime\prime}_{\cdot}$943). Both of the wavelength shift caused by the miscentering in slit and the atmosphere absorption are corrected by the comparison star. The detailed information of observations including observatory, instruments, and data reduction were described in F20.

\section{Results and Analysis}
Following the method in F20, we bin each spectrum with 50 \AA\ and fit the spectra by a power-law ($F_{\lambda} = A \lambda^{-\alpha}$). Then six bins are used for data analysis. The LC of each bin and spectral indices are shown in Figure 1. In this analysis, we use both of the data observed during 2018-2019 (provided in F20, hereafter "Epoch1") and 2019-2020 (hereafter "Epoch2"). The data of Epoch2 are listed in Table 1. We find that the brightness of S5 0716+714 during Epoch2 is much brighter than Epoch1 (see Figure 1), and there are several prominent flares during Epoch2. However, the amplitudes of variability from bin1 to bin6 during Epoch2 are 32.0\%, 31.0\%, 30.5\%, 30.2\%, 30.0\%, and 30.0\%, respectively, which are smaller than those in Epoch1. In this paper, we also test the time delay among different LCs. The right lower panel of Figure 1 gives an example of the interpolated cross-correlation function \citep[ICCF;][]{WP94} for all data. Although the sampling cadence and temporal span are high enough, we do not find any significant time lags among different bins. Thus, we verify that the optical variability of different wavelength should be generated in the same region and by the same radiation mechanism.

A similar variability behavior between spectral indices and LCs emerges in Figure 1. In order to investigate the spectral behavior, we plot the spectral index $\alpha$ vs. flux density of Bin1 $F_{\rm{\lambda}}$(Bin1) in Figure 2. There is a significant BWB trend in Figure 2, and this BWB behavior is qualitatively consistent with the predication of the shock-in-jet model \citep[e.g.][]{MG85}. A more interesting phenomenon is that the BWB trend is clearly dependent on the brightness. $\alpha$ increases roughly with the brightness during both Epoch1 (lower state) and Epoch2 (higher state), but a more significant BWB trend occurs at weaker state. There is a widely reported phenomenon in blazars that the synchrotron peak frequencies are well correlated with the peak luminosities \citep[e.g.,][]{Ma04b,Zh13}. F20 proposed that the BWB trends may depend on the relative locations of the synchrotron peak frequencies with respect to the observational frequency range. During different brightness states, the observational frequencies will cover different ranges with respect to the peak frequencies, and then different spectral indices will be detected. In order to test whether this scenario can reproduce the brightness-dependent BWB trends, synchrotron spectra with different $\nu_{\rm{p}}$ will be simulated and used to get the corresponding spectral indices and flux densities in the observational frequency range.

The synchrotron spectra are generated by a simple log-parabola function. Based on electron energy-dependent acceleration probability mechanisms (statistical acceleration mechanisms) \citep{Ma04a,Ma04b}, and/or on stochastic acceleration mechanisms \citep{Ka62,Tr11}, a log-parabolic spectral function can be produced for synchrotron emission of blazars. The log-parabolic law has been successfully applied to the synchrotron spectra within wide energy ranges for Mrk 421 \citep{Ma04a}, Mrk 501 \citep{Ma04b}, and \emph{Fermi} bright blazars \citep{Ch14}. In $\log(\nu f_{\nu}$) vs. $\log\nu$ plot, the model can be expressed as
\begin{equation}
  \log(\nu F_{\nu}) = \log(\nu_{\rm{p}} F_{\nu_{\rm{p}}}) -
  b \left[ \log\left(\frac{\nu}{\nu_{\rm{p}}}\right)\right]^{2},
\end{equation}
where $b$ is the spectral curvature, and $\nu_{\rm{p}}$ and $F_{\nu_{\rm{p}}}$ are peak frequency and flux density, respectively. The value of $b$, $\nu_{\rm{p}}$, and $F_{\nu_{\rm{p}}}$ can be obtained by fitting the SED of simultaneous wide frequency data. However, our optical data can not allow us to fit the SED of S5 0716+714. \citet{Fa16} fitted the SED of S5 0716+714 using the data corrected from NED (note that the data are not simultaneous), and provided the values of$b$, $\nu_{\rm{p}}$, and peak luminosity $L_{\rm{p}}$, which are 0.10, 10$^{14.96}$ Hz and 10$^{46}$ erg s$^{-1}$, respectively. Considering that $F_{\lambda} \propto \lambda^{-\alpha}$ are equivalent to $\nu f_{\nu} \propto \nu^{\alpha - 1}$ and the values of $\alpha$ are 0.23--1.14, most of our observations should locate at the right of the synchrotron peak frequencies (in $\log(\nu f_{\nu}$) vs. $\log \nu$ plot). This means $\nu_{\rm{p}} \le$10$^{14.7}$ Hz ($\sim$6000 $\mathring{\rm{A}}$) for our observations, and the brightness in \citet{Fa16} should be brighter than the brightness at Epoch2.

Figure 2 shows that most values of $\alpha$ are less than 1 (only one exception). The value of $\alpha$ in the brightest spectrum is close to 1. This means that the observational frequency range should just locate around the peak frequency of the brightest spectrum. Thus, we regard the brightest spectrum as the upper limit of ($F_{\nu_{\rm{p}}}$,$\nu_{\rm{p}}$). In fact, $L_{\rm{p}}$ and $\nu_{\rm{p}}$ follow a power-law relation, $L_{\rm{p}} \propto \nu_{\rm{p}}^{\alpha_{L}}$, in individual sources \citep[e.g.,][]{Ma08,Tr07,Tr09,Tr11,Zh13}. For the synchrotron peak, there are \citep[e.g.,][]{Ma08}:
\begin{equation}
    L_{\rm{p}} \propto N \gamma^2 B^2 \delta^4,
\end{equation}
and
\begin{equation}
  \nu_{\rm{p}} \propto \gamma^2 B \delta,
\end{equation}
where $N=\int N(\gamma)d \gamma$ is the total number of emitting electrons ($N(\gamma)$ is electron energy distribution [EED]), $\gamma m_{\rm{e}}c^2$ is the typical electron energy, $B$ is the magnetic field strength, and $\delta$ is the beaming factor \citep[see also][]{Ch18}. Formally, $\alpha_{L}$ = 1 applies as the spectral changes are dominated by variations in the electron average energy $\bar{\gamma}$, $\alpha_{L}$ = 2 applies as the spectral changes are dominated by changes of $B$, and $\alpha_{L}$ = 4 if changes in $\delta$ dominate. Special statement of $\alpha_{L}$ = 1 was presented in \citet{Ma08}. Equation (4) in \citet{Ch18} indicates that the expression of $L_{\rm{p}} \propto N(\gamma) \gamma^3 B^2 \delta^4$ is consistent with formulas (2).

Based on $\alpha_L$ and $b$, the synchrotron spectra can be rewritten as
\begin{equation}
\log(\nu F_{\nu})= \log A_{\rm{L}} +\alpha_L\log\nu_{\rm{p}}-b(\log\nu - \log\nu_{\rm{p}})^2,
\end{equation}
as $\nu_{\rm{p}} F_{\nu_{\rm{p}}} = A_{\rm{L}} \nu_{\rm{p}}^{\alpha_{L}}$ ($\equiv F_{\rm{p}}$). In fact, $b$ is a function of $\nu_{\rm{p}}$ \citep{Ma08,Tr07,Tr09,Tr11,Ch14,Ch18}, and $b\approx r/5$ with $r$ being the curvature of $N(\gamma)$ \citep{Ma04a}. Anti-correlations between $b$ and $\nu_{\rm{p}}$ were observed in high-energy peaked BL Lacs \citep{Ma08,Tr07,Tr09,Tr11}, and in \emph{Fermi} bright blazars \citep{Ch14,Ch18}. Their results indicate that $b$ roughly ranges from 0.15 to 0.6 in several TeV BL Lacs. $\log \nu_{\rm{p}}$ $\sim$ 15 for S5 0716+714, but for these TeV BL Lacs $\log \nu_{\rm{p}}$ $\sim$ 17. So, $b\approx$ 0.15--0.6 are taken as a reference for S5 0716+714. Furthermore, $b$ should be extended to $\sim$ 1 by the power-law best fit to ($\nu_{\rm{p}}$,$b$) of these TeV BL Lacs. Thus, we will adopt $b_{\rm{min}} = 0.2$ and $b_{\rm{max}} = 0.7$ for an assumed power-law of $b = A_{\rm{b}}\nu_{\rm{p}}^{-\alpha_{\rm{b}}}$ with $\alpha_{b}>0$. $\log \nu_{\rm{p}}(\rm{min}) = \log \nu_{\rm{p}}(\rm{max}) - \log 10/ \alpha_{\rm{L}}$ can be derived from assumption of $F_{\rm{p}}(\rm{max})=10 \times F_{\rm{p}}(\rm{min})$ for $\log \nu_{\rm{p}}(\rm{max})=14.7$ at the brightest spectrum and a given $\alpha_{\rm{L}}$. In fact, the variation ranges of $F_{\rm{p}}$ are about 10 times for most of BL Lacs in \citet{Ma08}. Moreover, $A_{\rm{L}}$ can be estimated for a given $\alpha_{\rm{L}}$ from $A_{\rm{L}} = F_{\rm{p}}(\rm{max}) \nu_{\rm{p}}^{-\alpha_{L}}(\rm{max})$. Also, $A_{\rm{b}}$ and $\alpha_{\rm{b}}$ can be obtained from $\nu_{\rm{p}}(\rm{min})$, $\nu_{\rm{p}}(\rm{max})$, $b_{\rm{min}}=0.2$, $b_{\rm{max}}=0.7$, and $b = A_{\rm{b}}\nu_{\rm{p}}^{-\alpha_{\rm{b}}}$. Thus, a series of $\nu_{\rm{p}}$ and artificial synchrotron spectrum can be generated from Equation (4) for a given $\alpha_{\rm{L}}$ to measure $\alpha$ and $F_{\rm{\lambda}}$(Bin1), which will be compared to the data distribution in Figure 2. Figure 3 shows an example of simulated spectra with $\alpha_{L}$ = 2. The measured $F_{\lambda}$(Bin1) and $\alpha$ are plotted in Figure 2 with color lines of $\alpha_{L}$ =1, 2, and 4. The simulation results are basically consistent with our observational data.

In general, the physical processes are different for short-term (ST) and long-term (LT) variability. Our monitoring campaign spans nearly two year with high cadence, and therefore, our observations include LT and ST variability. The $\alpha$--$F_{\lambda}$ correlation might be confused by the two different variability. The LT influence on the BWB significance can well be removed by the flux variation rate $\dot{F_{\lambda}}$ and the spectral index variation rate $\dot{\alpha}$ (F20). During several nights, we obtained more than one spectrum within a few hours. The small variability amplitude and high temporal resolution will generate extremely large error bars for $\dot{\alpha}$ and $\dot{F_{\lambda}}$. Hence, the variation rates calculated with intra-night observations are excluded from our analysis. The $\dot{\alpha}$--$\dot{F_{\lambda}}$(Bin1) distribution at Epoch1 is significantly different from that at Epoch2, and the best fitting is also significantly different from each other (see Figure 4). There are the same cases for the data of $\alpha$--$F_{\lambda}$(Bin) at Epoch1 and Epoch2 (see Figure 2). The BWB trend seems more significant at the lower state than at the higher state. The BWB trend seems to become saturated at the highest state (see Figure 2). We also test the relative variation rate of flux density $\dot{F_{\lambda}}/F_{\lambda}$ for Bin1. The distribution of $\dot{\alpha}$--$\dot{F_{\lambda}}/F_{\lambda}$(Bin1) at Epoch1 is basically consistent with that at Epoch2, and the best fitting to the data at Epoch1 is also roughly consistent with that at Epoch2 (see Figure 5). This indicates that the main variability mechanisms at different brightness states are the same as each other.

The Spearman's rank correlation test shows a positive correlation for $\alpha$--$F_{\rm{\lambda}}$(Bin1), and strong positive ones for $\dot{\alpha}$--$\dot{F_{\lambda}}$(Bin1) and $\dot{\alpha}$--$\dot{F_{\lambda}}/F_{\lambda}$(Bin1) (see Table 2). Same as in F20, a Monte Carlo (MC) simulation is used to reproduce these parameters and to confirm the Spearman's rank test results. For each pair of these parameters, each data array generated by the MC simulation is fitted with the SPEAR \citep{Pr92}, and the fitting gives the Spearman's rank correlation coefficient $r_{\rm{s}}$ and the p-value of hypothesis test $P_{\rm{s}}$. Assuming Gaussian distributions of X and Y, $r_{\rm{s}}$ and $P_{\rm{s}}$ distributions are generated by the SPEAR fitting to the data of X and Y from $10^4$ realizations of the MC simulation. Averages, $r_{\rm{s}}$(MC) and $P_{\rm{s}}$(MC), are calculated by the $r_{\rm{s}}$ and $P_{\rm{s}}$ distributions, respectively (see Table 2). Standard deviations of distributions are taken as the relevant uncertainties. The MC simulation results confirm the ordinary Spearman's rank test results. Thus, these correlations for $\alpha$--$F_{\lambda}$(Bin1), $\dot{\alpha}$--$\dot{F_{\lambda}}$(Bin1), and $\dot{\alpha}$--$\dot{F_{\lambda}}/F_{\lambda}$(Bin1) will be reliable. In fact, the significant levels of $\alpha$--$F_{\lambda}$(Bin) correlations are lower than those of $\dot{\alpha}$--$\dot{F_{\lambda}}$(Bin1) ones, and $\dot{\alpha}$--$\dot{F_{\lambda}}/F_{\lambda}$(Bin1) ones, at Epoch1 and Epoch2, respectively (see Table 2). This confirms that the LT influence on the BWB significance can well be removed by $\dot{F_{\lambda}}$ and $\dot{\alpha}$ (F20).

\section{Discussion and Conclusions}

Although various spectral behaviors of S5 0716+714 have been widely reported, rare studies consider the effect of the brightness state (see Section 1) on the behaviors. Generally, the BWB trends are the most accepted phenomenon in jet dominated BL Lacs while other chromatic behaviors are still controversial. During our monitoring program, the source showed a strong BWB trend (see Figure 2), but the correlation between brightness and spectral index becomes weaker toward the brighter state. The brightness-dependent BWB trend is only significant when the variability covers a large brightness range ($>$ 2 mag) and the data sampling is high enough. This usually requires long-term continuous monitoring programs. In fact, the cadence of all previous long-term observations are much lower than ours, and the large gaps in the color vs. magnitude plots will hide the detailed effect of brightness on the spectra. While other short-term observations can not provide sufficient brightness coverage. Hence, the different index-flux correlations might be caused by the different brightness states.

Several models are proposed to interpret the BWB chromatism in BL Lacs. The models based on the contribution of accretion disc \citep[e.g.,][]{Ra07} or host galaxy \citep[e.g.,][]{Ni08} can be excluded for the jet-dominated sources. The shock-in-jet model is the most favorable interpretation, where the change of $\alpha$ is related to the electron acceleration mechanism in a relativistic shock propagating down a jet \citep{MG85}. The electron average energy and the magnetic field in the pre-shock region can be amplified by the shock. The variations of their values will shift $\nu_{\rm{p}}$, and then generate different spectral index in the observational window. In addition, the turbulence in the post-shock region will emerge in the relativistic jet, and can strongly amplify the magnetic field $B$ in the post-shock region \citep[e.g.,][and references therein]{Mi14}. The turbulence amplification of $B$ will make $\nu_{\rm{p}}$ to higher values. The higher $B$ may make $\nu_{\rm{p}}$ to the right of the observational frequency range, if formerly $\nu_{\rm{p}}$ is just next to the left of the observational frequency range. This turbulence process might produce the special data point around $\alpha \sim 1.2$ in Figure 2.

In fact, $\alpha_{\rm{L}}$ may be smaller than 1 derived from observations, e.g., $\alpha_{\rm{L}}\approx$ 0.4 for Mrk 421 \citep{Tr09}. For $\alpha_{\rm{L}}=$ 0.4, we also simulated the synchrotron spectra with Equation (4), and measured $F_{\lambda}$(Bin1) and $\alpha$ in the observational window. The derived curve is roughly consistent with these data of the higher state (see Figure 2). There will be different values of $\alpha_{\rm{L}}$ for the same object, because there may be a broken distribution of $L_{\rm{p}}$ and $\nu_{\rm{p}}$, e.g., Mrk 421 \citep{Tr09}. Formally, it is reasonable for different values of $\alpha_{\rm{L}}$ for the same object. Thus, there might be smaller values of $\alpha_{\rm{L}}$ for S5 0716+714, and this possibility needs to be tested with radio, IR--optical--UV, and soft X-ray observations. The derived curves of $\alpha_{\rm{L}} =$0.4, 1.0, 2.0, and 4.0 as a whole are well consistent with these observational data of Epoch1 and Epoch 2 (see Figure 2). The derived curves of $\alpha =$ 1.0 and 2.0 are basically consistent with these observational data. Thus, the optical variability in our observations of S5 0716+714 is likely dominated by the variations of the EED and the magnetic field in the emitting region.

\citet{Ch14} obtained a significant anti-correlation between $b$ and $\nu_{\rm{p}}$ of the synchrotron component for \emph{Fermi} bright blazars, and the slope of the correlation is consistent with the prediction of a stochastic acceleration scenario. \citet{Ma08} found a significant anti-correlation between $b$ and $\nu_{\rm{p}}$ for TeV BL Lacs, which points toward statistical/stochastic acceleration processes for the emitting electrons. Other researches also found  anti-correlations between $b$ and $\nu_{\rm{p}}$ \citep{Tr07,Tr09,Tr11}. Recently, \citet{An20} found that BL Lacs show an anti-correlation between $r$ of a log-parabolic EED and its peak energy $\gamma_{\rm{p}}$, which is a signature of stochastic acceleration. Particle acceleration mechanisms can produce the log-parabolic EED \citep[see e.g.,][]{Ka62,Ma04a,Ma06,Tr11,Ch14}. The intrinsic curvature $r$ in the EED arises due to the combined effect of particle acceleration and radiative cooling \citep[e.g.,][]{An20}. The anti-correlation between $b$ and $\nu_{\rm{p}}$ might originate from that between $r$ and $\gamma_{\rm{p}}$. Thus, our simulation of the log-parabolic synchrotron spectra can well reproduce the $\alpha$--$F_{\lambda}$(Bin1) distribution trend in Figure 2. In addition, Figure 3 shows that the observed BWB trend in Figure 2 is likely controlled by the relative position changes of $\nu_{\rm{p}}$ with respect to the observational window.

The simulation results in Figure 2 show that $\alpha_{L}$ = 4 is only just on the edge of the observational data. Therefore, the brightness-dependent BWB trend should not be dominated by the changes in $\delta$, i.e., the observed BWB trend in Figure 2 should not be from the changes of the bulk velocity and/or the viewing angle of the emitting region in jet. \citet{Ra03} found that the long-term variation of $\delta$ reaches a factor of $\sim$ 1.3. \citet{Zh12} fitted the SEDs of 10 BL Lacs (including S5 0716+714) in both low and high states. For a certain source, the value of $\delta$ is basically constant, and the variation of $B$ is larger than that of $\delta$ (see their Table 1). Furthermore, the relative position changes of $\nu_{\rm{p}}$ with the observational window can significantly influence the BWB trend. Thus, the data distribution in the $\alpha$--$F_{\lambda}$(Bin1) diagram may be dominated by the variations of $B$ and $\bar{\gamma}$ rather than that of $\delta$ for our observations of S5 0716+714. Figure 3 shows that the brightness-dependent BWB trend is a natural result of the synchrotron spectrum changes calculated from Equation (4) and $b \propto \nu_{\rm{p}}^{-\alpha_{\rm{b}}}$. The shifts of $\nu_{\rm{p}}$ and $L_{\rm{p}}$ will nonlinearly change $\alpha$ in the $\alpha$-$F_{\lambda}$(Bin1) diagram. Though, the curve of $B$-dominated spectral changes is above the one of $\bar{\gamma}$-dominated spectral changes in Figure 2, we can not determine the $B$-dominated region and the $\bar{\gamma}$-dominated region, because of possible combinations of $B$ and $\bar{\gamma}$ contributions. In fact, the relevant parameters are likely changing at the same time.

The new observational data reveal an extremely bright state with $\alpha = 1$ at Epoch2, and it seems that the BWB trend becomes saturated at the highest state at Epoch2. Combined with the observational data of Epoch1, we find a significantly brightness-dependent BWB chromatism (see Figure 2). The shock-in-jet model predicts the BWB behaviors for blazars. Our observations during Epoch1 and/or Epoch2 confirm in principle its prediction, which lacks details of the BWB data distribution, such as that in Figure 2. The observed BWB trend roughly becomes weaker towards the brighter state. On average, the brightness at Epoch2 is higher than that at Epoch1. The BWB trend at Epoch1 differs significantly from that at Epoch2 in the diagrams of $\alpha$--$F_{\lambda}$(Bin1) and $\dot{\alpha}$--$\dot{F_{\lambda}}$(Bin1), and the best fittings are the same cases (see Figures 2 and 4). The $\dot{\alpha}$--$\dot{F_{\lambda}}/F_{\lambda}$(Bin1) distribution at Epoch1 is basically consistent with that at Epoch2, and the best fitting is the same case (see Figure 5). This indicates that the main variability mechanisms at the different brightness states are the same as each other. A special value of $\alpha \sim 1.2$ in Figure 2 may result from the magnetic field amplification due to the turbulence generated in the post-shock region. The simulated synchrotron SED variability well reproduces the brightness-dependent BWB trend in Figure 2. The BWB trend is seemingly dominated by the relative position changes of $\nu_{\rm{p}}$ with respect to the observational frequency range, and effectively may be controlled by the variations of $\bar{\gamma}$ and $B$ in the emitting region.

\acknowledgements {We are grateful to the anonymous referee for constructive comments leading to significant improvement of this paper. We thank the financial support of the Key Research Program of the Chinese Academy of Sciences (CAS: grant No. KJZD-EWM06), the National Natural Science Foundation of China (NSFC; grants No. 11433004, 11991051, 11703077, 11573067, and 11973050), and the Ministry of Science and Technology of China (2016YFA0400700). We are also thankful for the joint fund of Astronomy of the NSFC and the CAS (grants No. U1831125, U1331118, and U1831135), and the CAS Interdisciplinary Innovation Team. We acknowledge the support of the staff of the Lijiang 2.4 m telescope. Funding for the telescope has been provided by Chinese Academy of Sciences and the People's Government of Yunnan Province. }

\clearpage

\begin{deluxetable}{ccccccc}
  \tablecolumns{7}
  \setlength{\tabcolsep}{5pt}
  \tablewidth{0pc}
  \tablecaption{Spectral flux in each bin for Epoch 2}
  \tabletypesize{\scriptsize}
  \tablehead{
  \colhead{JD}                        &
  \colhead{Bin1}                      &
  \colhead{Bin2}                      &
  \colhead{Bin3}                      &
  \colhead{Bin4}                      &
  \colhead{Bin5}                      &
  \colhead{Bin6}
} \startdata
2458756.428773 & 1.155 $\pm$ 0.038 & 1.038 $\pm$ 0.024 & 0.954 $\pm$ 0.017 & 0.863 $\pm$ 0.018 & 0.840 $\pm$ 0.019 & 0.753 $\pm$ 0.020 \\
2458770.326829 & 2.574 $\pm$ 0.046 & 2.264 $\pm$ 0.021 & 2.039 $\pm$ 0.022 & 1.804 $\pm$ 0.021 & 1.678 $\pm$ 0.026 & 1.502 $\pm$ 0.024 \\
2458775.328299 & 1.771 $\pm$ 0.032 & 1.569 $\pm$ 0.036 & 1.408 $\pm$ 0.028 & 1.283 $\pm$ 0.022 & 1.208 $\pm$ 0.020 & 1.091 $\pm$ 0.020 \\
2458776.387940 & 1.738 $\pm$ 0.038 & 1.532 $\pm$ 0.021 & 1.398 $\pm$ 0.016 & 1.264 $\pm$ 0.023 & 1.195 $\pm$ 0.023 & 1.073 $\pm$ 0.017 \\
2458777.336424 & 1.893 $\pm$ 0.043 & 1.646 $\pm$ 0.022 & 1.503 $\pm$ 0.025 & 1.332 $\pm$ 0.024 & 1.261 $\pm$ 0.024 & 1.129 $\pm$ 0.021 \\
... & ... & ... & ... & ... & ... & ... \\
\enddata
\tablecomments{\footnotesize This table is available in its entirety in machine-readable form.}
\label{Table1}
\end{deluxetable}

\begin{deluxetable}{cccccc}
  \tablecolumns{6}
  \setlength{\tabcolsep}{5pt}
  \tablewidth{0pc}
  \tablecaption{Spearman's rank analysis results for Epoch1 and Epoch 2}
  \tabletypesize{\scriptsize}
  \tablehead{\colhead{X}  &  \colhead{Y} &  \colhead{$r_{\rm{s}}$} &  \colhead{$P_{\rm{s}}$}
  &  \colhead{$r_{\rm{s}}$(MC)} &  \colhead{$-\log P_{\rm{s}}$(MC)}
} \startdata

$\dot{\alpha}$ & $\dot{F_{\lambda}}^{\dag}$ & 0.800 &  $< 10^{-4}$ &  0.70$\pm$0.05  &  7.3$\pm$1.5  \\

 $\dot{\alpha}$ & $\dot{F_{\lambda}}^{\ddag}$ & 0.719 &  $< 10^{-4}$ &  0.62$\pm$0.05  &  11.3$\pm$2.0  \\

$\dot{\alpha}$ &$\dot{F_{\lambda}}/F_{\lambda}^{\dag}$ & 0.787  & $< 10^{-4}$  & 0.69$\pm$0.06 & 7.2$\pm$1.5  \\

$\dot{\alpha}$ &$\dot{F_{\lambda}}/F_{\lambda}^{\ddag}$ & 0.736  & $< 10^{-4}$  & 0.63$\pm$0.04 & 12.1$\pm$2.0  \\

$\alpha$ &$F_{\lambda}^{\dag}$ & 0.544  & $< 10^{-4}$  & 0.53$\pm$0.03 & 4.0$\pm$0.4  \\

$\alpha$ &$F_{\lambda}^{\ddag}$ & 0.407  & $< 10^{-4}$  & 0.39$\pm$0.03 & 4.4$\pm$0.7  \\

\enddata
\tablecomments{\footnotesize X and Y are the relevant quantities presented in Figures 2, 4, and 5. $\dag$ and $\ddag$ denote the regression analyses of the data at Epoch1 and Epoch2, respectively.}
\label{Table8}
\end{deluxetable}

\begin{figure*}
  \includegraphics[scale = 0.8]{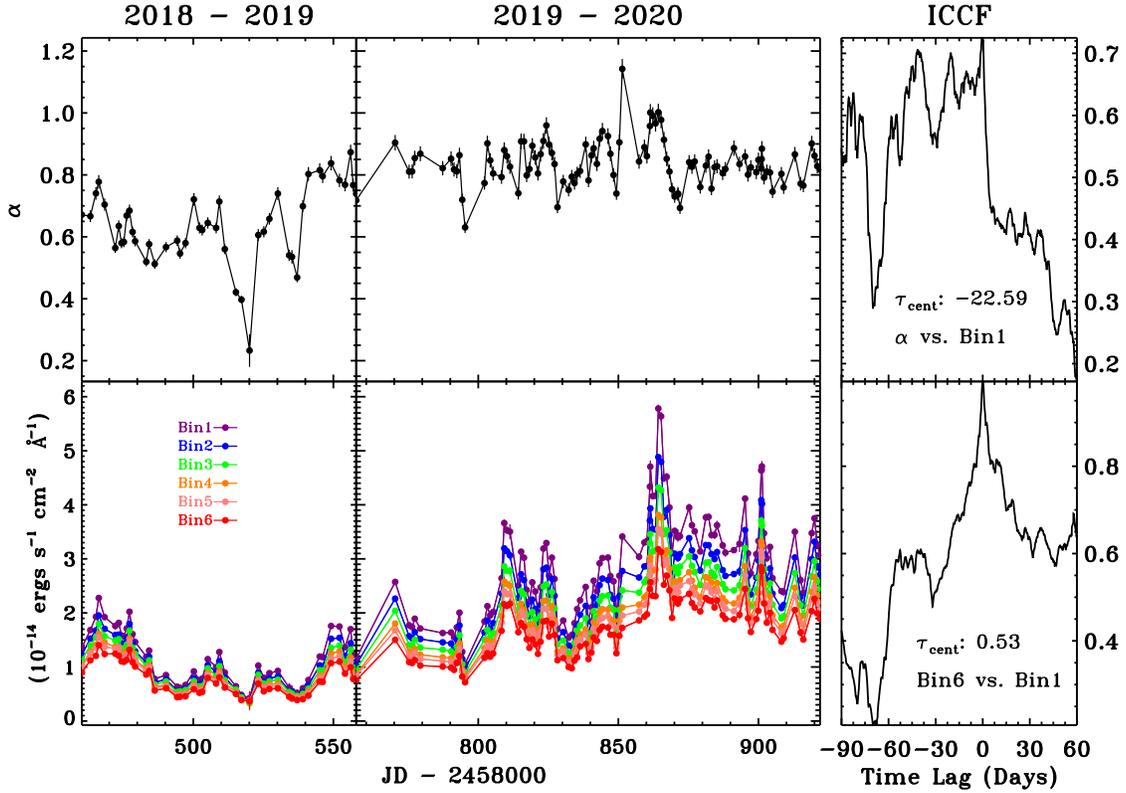}
 \caption{The left panel is spectral indices (top) and light curves (bottom). The right panel is interpolation cross-correlation functions.}
  \label{fig1}
\end{figure*}

\clearpage

\begin{figure*}
 \begin{center}
  \includegraphics[angle=0,scale=0.8]{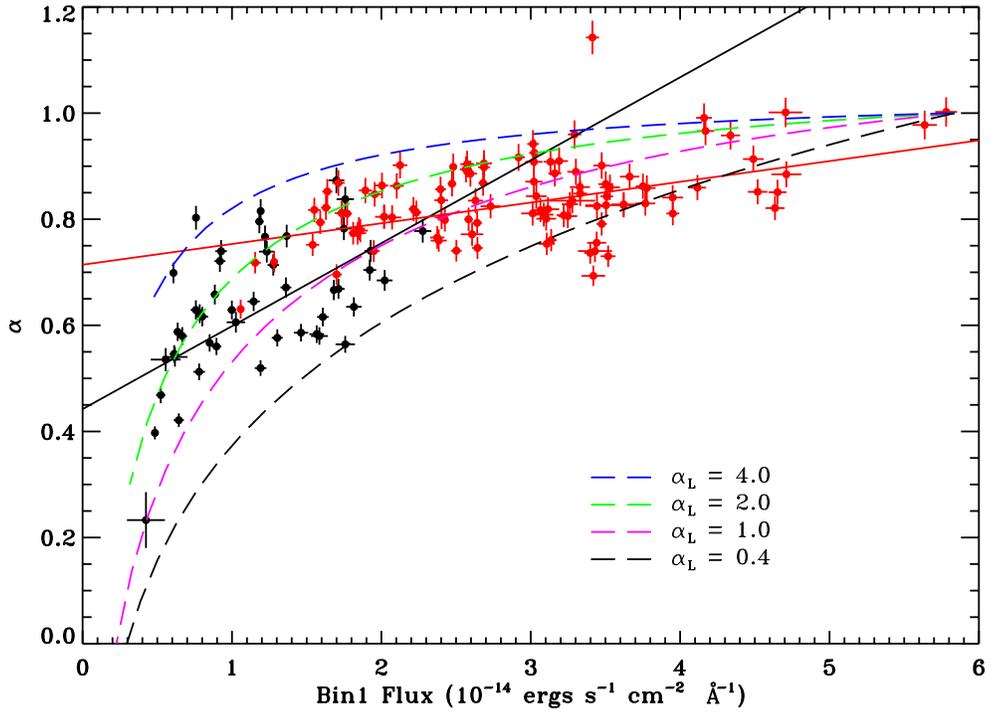}
 \end{center}
 \caption{Correlations between $\alpha$ and $F_{\rm{\lambda}}$(Bin1). The black and red dots denote the data during Epoch1 and Epoch2, respectively. The dashed lines are simulation results. The solid lines are the best fitting to data points by the FITEXY estimator \citep{Pr92}: $y=0.44+0.16\times x$ for Epoch1 (the black solid line) and $y=0.71+0.04\times x$ for Epoch2 (the red solid line).}
  \label{fig2}
\end{figure*}

\clearpage

\begin{figure*}
 \begin{center}
  \includegraphics[angle=0,scale=0.8]{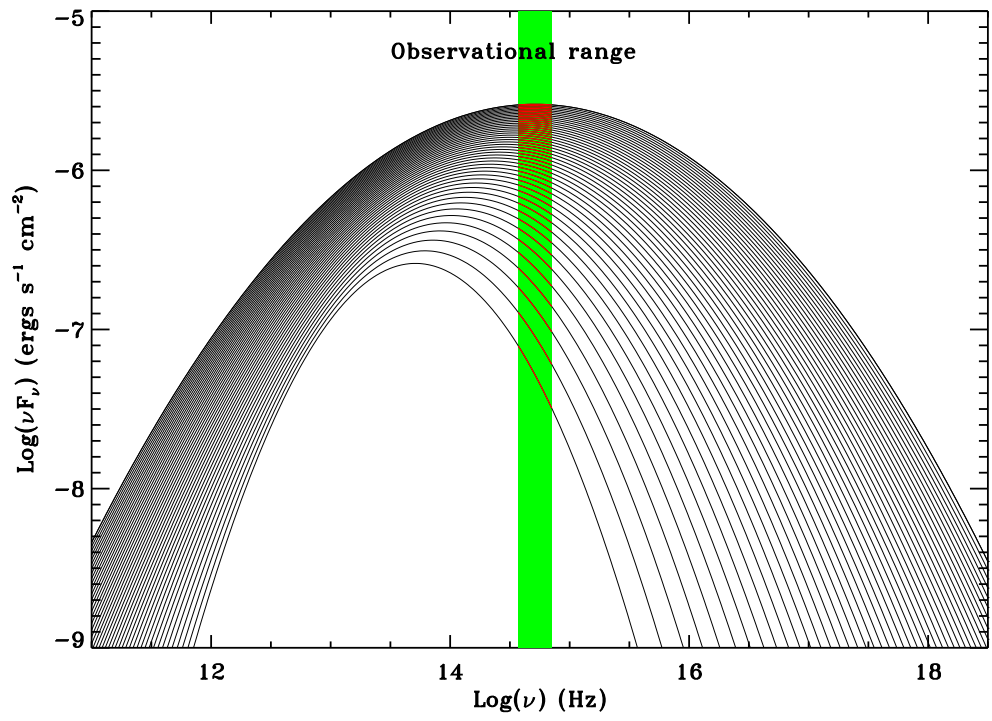}
 \end{center}
 \caption{The synchrotron spectra generated by $\log$-parabola function.}
  \label{fig3}
\end{figure*}

\begin{figure*}
 \begin{center}
  \includegraphics[angle=-90,scale=0.45]{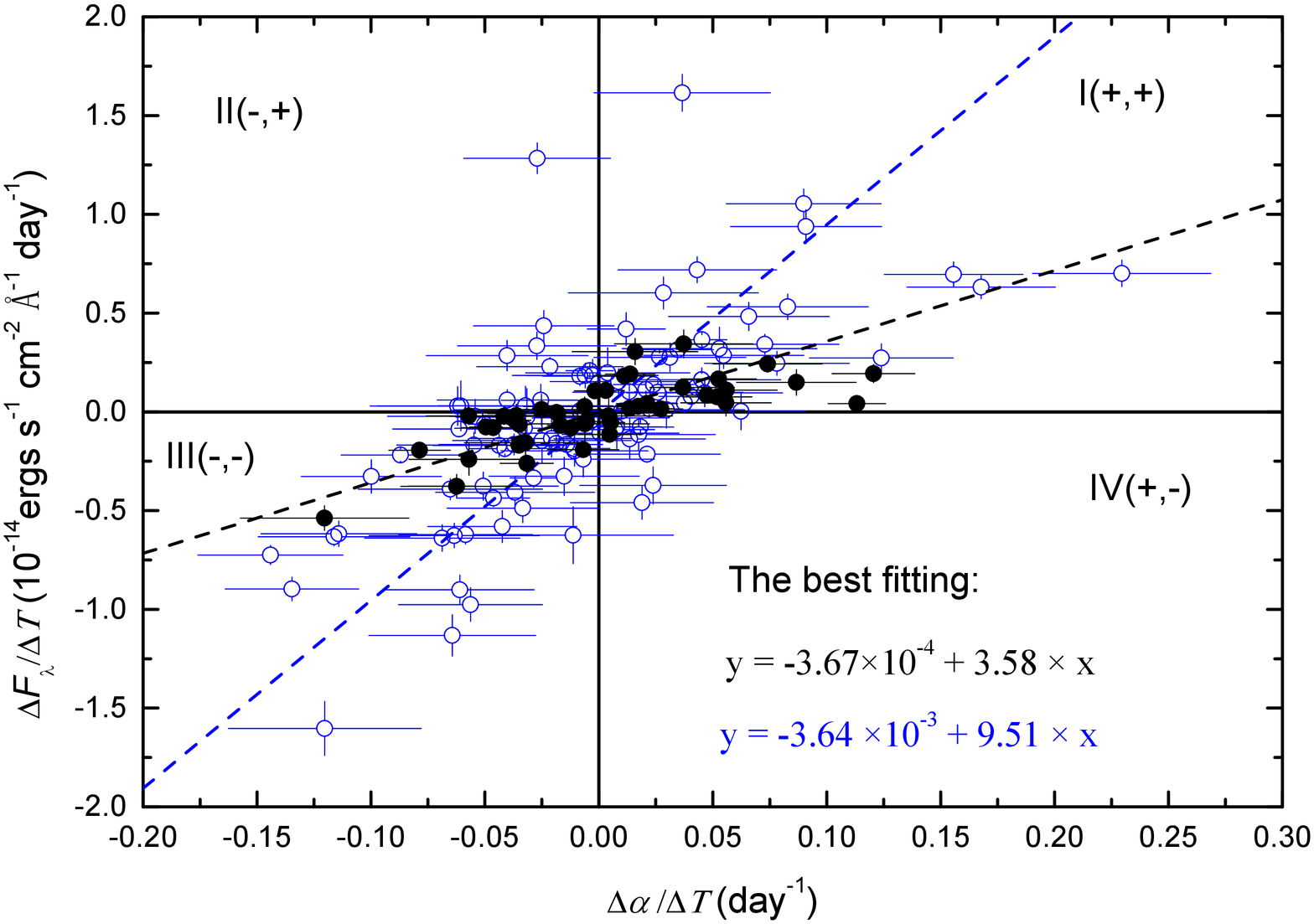}
 \end{center}
 \caption{$\dot{F_{\lambda}}$(Bin1) vs. $\dot{\alpha}$. The black dots and blue circles denote the data during Epoch1 and Epoch2, respectively. The FITEXY estimator \citep{Pr92} gives the best linear fitting (the dashed lines).}
  \label{fig4}
\end{figure*}

\begin{figure*}
 \begin{center}
  \includegraphics[angle=-90,scale=0.45]{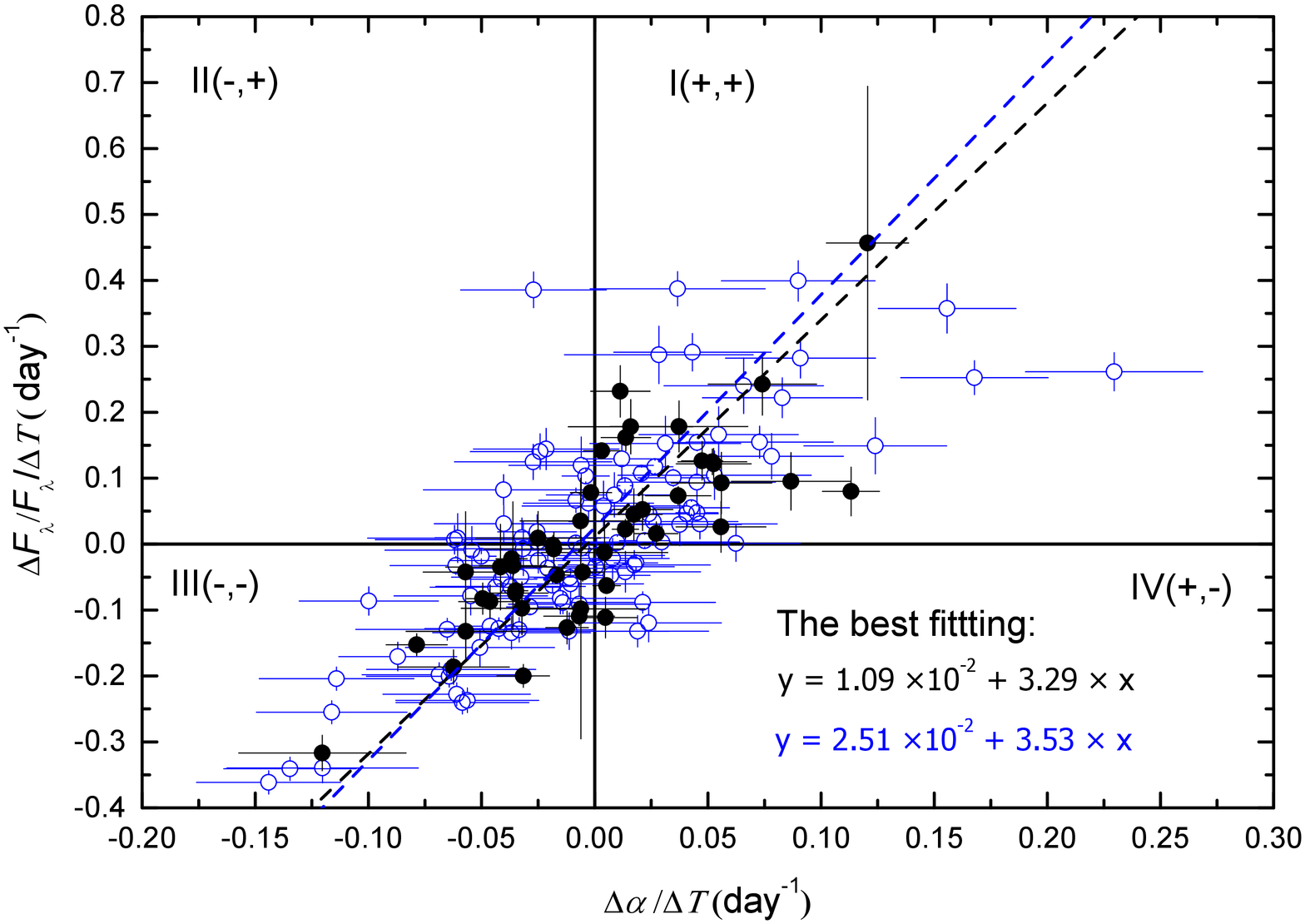}
 \end{center}
 \caption{$\dot{F_{\lambda}}/F_{\lambda}$(Bin1) vs. $\dot{\alpha}$. The black dots and blue circles denote the data during Epoch1 and Epoch2, respectively. The FITEXY estimator gives the best linear fitting (the dashed lines).}
  \label{fig4}
\end{figure*}

\end{document}